\documentclass[twocolumn,showpacs,floatfix,superscriptaddress,amsmath,amssymb,prl]{revtex4}
\usepackage{amsfonts}
\usepackage{bbm}
\usepackage{mathrsfs}
\usepackage{txfonts}
\usepackage{amssymb}
\usepackage{graphicx}
\usepackage{hyperref}
\usepackage{ulem}
 \usepackage{overpic}
 \usepackage{psfrag}
\usepackage{tabularx}
\usepackage{array}
\usepackage{placeins}
\makeatletter
\renewcommand\subsection{\@startsection
{subsection}{2}{0mm}
 {-\baselineskip}
 {0.5\baselineskip}
{\FloatBarrier\normalfont\Large\bfseries}}
\makeatother
\newcommand{\be}{\begin{equation}}
\newcommand{\ee}{\end{equation}}

\newcommand{\PreserveBackslash}[1]{\let\temp=\\#1\let\\=\temp}

\begin{document}
\title{Pairing mechanism for high temperature superconductivity in the cuprates: what can we learn from the two-dimensional $t-J$ model?}

\author{Huan-Qiang Zhou}
\affiliation{Centre for Modern Physics and Department of Physics,
Chongqing University, Chongqing 400044, The People's Republic of
China}

\begin{abstract}
More than twenty years have passed since high temperature
superconductivity in the copper oxides (cuprates) was discovered by
J.G. Bednorz and K.A. M\"uller in 1986~\cite{muller}. Although
intense theoretical and experimental efforts have been devoted to
the investigation of this fascinating class of  materials, the
pairing mechanism responsible for unprecedented high transition
temperatures $T_c$ remains elusive. Theoretically, the difficulty
lies in the fact that this class of materials, as doped Mott-Hubbard
insulators~\cite{anderson}, involve strong electronic correlations,
which renders conventional theoretical approaches  unreliable.
Recent progress in numerical simulations of strongly correlated
electron systems in the context of tensor network
representations~\cite{jordan,shi} makes it possible to get access to
information encoded in the ground-state wave functions of the
two-dimensional $t-J$ model-a minimal model, as widely believed, to
understand electronic properties of doped Mott-Hubbard
insulators~\cite{zhangrice,zhangetal,wenlee,leenagaosawen}. In this
regard, an intriguing question is whether or not the two-dimensional
$t-J$ model holds the key to understanding high temperature
superconductivity in the cuprates.  As it turns out, such a key lies
in a superconducting state with mixed spin-singlet $d+s-$wave and
spin-triplet $p_x (p_y)$-wave symmetries in the presence of an
anti-ferromagnetic background~\cite{li}. Here, the $d+s$-wave
component in the spin-singlet channel breaks $U(1)$ symmetry in the
charge sector, whereas both the anti-ferromagnetic order and the
spin-triplet $p_x (p_y)$-wave component breaks $SU(2)$ symmetry in
the spin sector. Therefore, four gapless Goldstone modes occur.
However, even if we resort to the Kosterlitz-Thouless
transition~\cite{kosterlitz}, {\it only} the $d+s$-wave
superconducting component survives thermal fluctuations. This turns
three gapless Goldstone modes, arising from $SU(2)$ symmetry
breaking, into two degenerate soft modes, with twice the
spin-triplet $p_x (p_y)$-wave superconducting energy gap as their
characteristic energy scale: one is a spin-triplet mode observed as
a spin resonance mode in inelastic neutron scattering, the other is
a spin-singlet mode observed as a $A_{1g}$ peak in electronic Raman
scattering. The scenario allows us to predict that pairing is of
$d+s$-wave symmetry, with the two degenerate soft modes as the
long-sought key ingredients in determining the transition
temperature $T_c$, thus offering a possible way to resolve the
controversy regarding the elusive mechanism for high temperature
superconductivity in the cuprates.

\end{abstract}
 \maketitle

\begin{figure}%
\centering
\begin{overpic}
 [width=0.45\textwidth,totalheight=30mm]{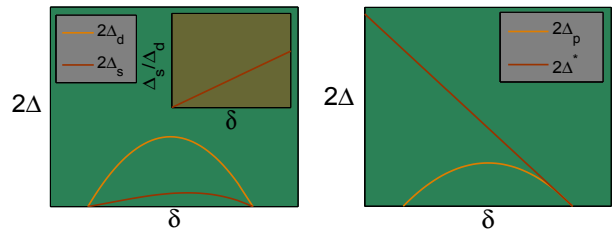}
  \end{overpic}
\caption{(color online) The doping dependence of the order parameters for a model system
describing doped Mott-Hubbard insulators on a two-dimensional square
lattice, whose ground-state wave function is a superconducting state
with mixed spin-singlet $d+s-$wave and spin-triplet $p_x (p_y)-$wave
symmetries in the presence of an anti-ferromagnetic background, with
the order parameters for the $s$-wave, $d$-wave, and $p_x
(p_y)$-wave superconducting components, together with the
anti-ferromagnetic order parameter $N$. Left panel: The superconducting gaps $\Delta_d$ and $\Delta_s$
for the spin-singlet $d$-wave and $s$-wave components as a function of doping $\delta$, with their ratio $\Delta_s/\Delta_d$ shown in the inset.
Right panel: The energy scales $2 \Delta^* \sim k_B T^* \sim N$ and $E_{\rm res} = 2 \Delta_{p}$ for the anti-ferromagnetic order and
the spin-triplet $p_x (p_y)-$wave superconducting order as a function of doping $\delta$, with $N$ being the anti-ferromagnetic order, and $k_B$
the Boltzmann constant. A crossover from the Bose-Einstein condensation (BEC) regime to the Bardeen-Cooper-Schrieffer (BCS) regime occurs,
when the two energy scales merge into one single energy scale in the (heavily) overdoped regime. Note that, in general, $E_{\rm res} < 2 \Delta_d$.
Indeed, $E_{\rm res} \approx 1.25 \Delta_d$, as predicted from the two-dimensional $t-J$ model.
In addition,  $E_{\rm res}$ scales with the superconducting transition
temperature $T_c$: $E_{\rm res} \sim k_B T_c$} \label{FIG1}
\end{figure}

Imagine if we would have been able to solve a model system
describing doped Mott-Hubbard insulators on a two-dimensional square
lattice, whose ground-state wave function is a superconducting state
with mixed spin-singlet $d+s-$wave and spin-triplet $p_x (p_y)-$wave
symmetries in the presence of an anti-ferromagnetic background, with
the order parameters for the $s$-wave, $d$-wave, and $p_x
(p_y)$-wave superconducting components, together with the
anti-ferromagnetic order parameter, shown in Fig.~\ref{FIG1}, in a
proper doping range. Note that $\Delta_d$ and $\Delta_s$ are,
respectively, the spin-singlet $d$-wave and $s$-wave superconducting
energy gaps, whereas $\Delta_p$ is the spin-triplet $p_x (p_y)$-wave
superconducting energy gap and  $N$ is the anti-ferromagnetic N\'eel
order parameter. A few peculiar features of this state are: (i) Both
the 90 degree (four-fold) rotation symmetry and the translation
symmetry under one-site shifts are spontaneously broken on the
square lattice. (ii) Spin-rotation symmetry $SU(2)$ is spontaneously
broken, due to the simultaneous occurrence of both the $p_x
(p_y)-$wave superconducting component and the anti-ferromagnetic
order. (iii) $U(1)$ symmetry in the charge sector is spontaneously
broken, due to pairing in both spin-singlet and spin-triplet
channels. Here, we emphasize that the symmetry mixing of the
spin-singlet and spin-triplet channels arises from the spin-rotation
symmetry breaking, {\it simply because spin is not a good quantum
number}. (iv) All superconducting components are homogeneous, in the
sense that their superconducting order parameters are independent of
sites on the lattice.

Now let us switch on thermal fluctuations. Suppose we restrict
ourselves to a strict two dimensional system. Then, even if the
Kosterlitz-Thouless transition~\cite{kosterlitz} is invoked, {\it
only} spin-singlet $d+s$-wave superconducting component survives
thermal fluctuations. However, the non-abelian $SU(2)$ symmetry is
not allowed to be broken at any finite
temperature~\cite{hohenberg,merminwagner}. This immediately implies
that the Goldstone modes arising from the spontaneous symmetry
breaking of $SU(2)$ in the spin sector have to be turned into
degenerate soft modes, with twice the spin-triplet $p_x (p_y)$-wave
superconducting energy gap as their characteristic energy scale: one
is a spin-triplet mode associated with the anti-ferromagnetic order,
with the momentum transfer $(\pi, \pi)$, and the other is a
spin-singlet mode associated with the spin-triplet $p_x (p_y)$-wave
superconducting component, with the momentum transfer $(0,0)$. On
the other hand, there is nothing to prevent from the breaking of the
discrete four-fold rotation symmetry on the square lattice.
Actually, this broken symmetry not only manifests itself in the
admixture of a small $s$-wave component to the dominant $d$-wave
superconducting state (see Fig.~\ref{FIG1}, left panel), but also
protects the spin-singlet soft mode that is unidirectional as it
arises from the $p_x (p_y)$-wave superconducting component.

Our argument leads to a scenario that, at any finite temperature,
the pairing is of $d+s$-wave symmetry, with two degenerate soft
modes acting as the key ingredients in determining the transition
temperature $T_c$. Actually, two distinct energy scales $2 \Delta^*$
and $E_{\rm res}$ are involved, in a marked contrast with the
conventional superconductors: $2 \Delta^*$ arises from the
anti-ferromagnetic N\'eel order parameter $N$, which is responsible
for pairing, with its coupling strength decreasing almost linearly
with doping, whereas $E_{\rm res} = 2 \Delta_p$, which is
responsible for condensation. Therefore,  $E_{\rm res}$ must scale
with the superconducting transition temperature $T_c$, i.e., $E_{\rm
res} \sim k_B T_c$, with $k_B$ being the Boltzmann constant [see
Fig.~\ref{FIG1}, right panel]. Similarly, $2 \Delta^*$ scales as $2
\Delta^* \sim k_B T^*$, with $T^*$ being the so-called pseudogap
temperature~\cite{ding,loeser}. In addition, one may expect that
$E_{\rm res} < 2 \Delta_d$, simply due to the fact that the
predominant $d$-wave superconducting component survives thermal
fluctuations. Considering that both the superconducting gap
$\Delta_d$ and the transition temperature $T_c$ characterize the
superconductivity, they should track each other in the entire doping
range, implying $\Delta_d \sim k_B T_c$. In fact, for the $t-J$
model, our simulation indicates that $E_{\rm res} \approx 1.25
\Delta_d$~\cite{li}. This in turn allows us to estimate a universal
coefficient $\kappa \approx 5.37$ in the scaling relation: $E_{\rm
res} = \kappa \; k_B T_c$.

Note that the two distinct energy scales in the underdoped regime
are split off from one single energy scale in the (heavily)
overdoped regime. This naturally results in a crossover from the
Bose-Einstein condensation (BEC) regime to the
Bardeen-Cooper-Schrieffer (BCS) regime, as conjectured in
Ref.~\cite{uemura}, which in turn is essentially equivalent to the
phase fluctuation picture proposed by Emery and
Kivelson~\cite{emery}. However, there is an important difference:
the superconductivity weakens in the heavily underdoped regime, not
only because of the loss of phase coherence, but also because of the
decrease of the superconducting gap $\Delta_d$ with underdoping.

Now a fundamental question is whether or not such a scenario is
really relevant to the high $T_c$ problem. This brings us to the
phenomenology of the high temperature cuprate superconductors.

\begin{figure}[t]%
\begin{overpic}
 [width=0.45\textwidth,totalheight=60mm]{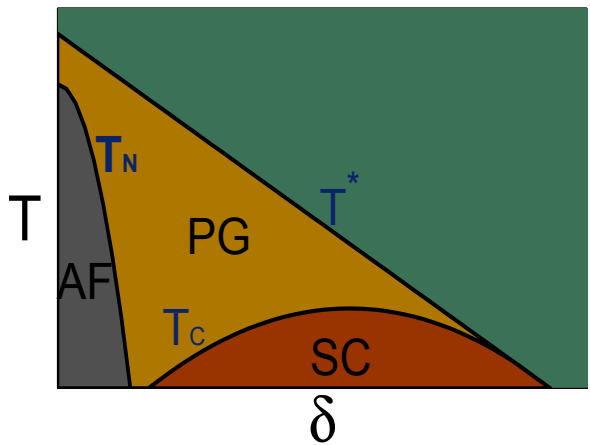}
  \end{overpic}
\caption{(color online)  A schematic phase diagram of the hole-doped high temperature cuprate superconductors plotted as a function of temperature $T$
and hole doping $\delta$. Here, SC, PG, and AF stand for the superconducting, pseudogap, and anti-ferromagnetic phases,
respectively.  $T_c$, $T^*$, and $T_N$ represent, respectively, the superconducting transition temperature,
the pseudogap temperature, and the anti-ferromagnetic N\'eel temperature.
In our scenario, the PG phase is characterized by incoherent
preformed pairs, with a pseudogap opening near the anti-nodal region, leaving the remnant gapless Fermi arcs near the nodes in the momentum space.
In addition, the four-fold rotation symmetry and the translation symmetry are broken on the square lattice.
Superconducivity occurs, when long range phase coherence develops at $T_c$.} \label{FIG2}
\end{figure}

First, let us focus on the two distinct energy scales $2 \Delta^*$
and $E_{\rm res}$.  Physically, the two distinct energy scales
measure, respectively, the pairing strength and the coherence of the
superfluid condensate. This naturally leads to two different phases:
one is characterized by incoherent pairing, which may be identified
with the pseudogap phase; the other is associated with the emergence
of a coherent condensate of superconducting pairs, which may be
identified with the superconducting phase of $d+s$-wave symmetry
[see Fig.~\ref{FIG2}]. Evidence for the two distinct energy scales
was reported in angle-resolved photoemission
spectra~\cite{tanaka,kondo,lee,yoshida}, electronic Raman
spectra~\cite{gallais,letacon,guyard1,guyard2}, scanning tunneling
microscopy~\cite{boyer}, $c$-axis conductivity~\cite{liyu}, Andreev
reflection~\cite{deustcher}, magnetic penetration
depth~\cite{xiang}, and other probes (for a review, see,
Ref.~\cite{hufner}. Actually, these studies indicate that the gap
near the antinodal region, which is identified as the pseudogap,
does not scale with $T_c$ in the underdoped regime, whereas the gap
near the nodal region may be identified as the superconducting order
parameter in the
cuprates~\cite{tanaka,kondo,lee,guyard1,yoshida,gallais,letacon}.
This identification offers a natural explanation why the two
distinct energy scales in the underdoped regime merge into one
single energy scale in the (heavily) overdoped regime as a
consequence of the evolution of the Fermi arcs in the underdoped
regime to a large Fermi surface in the (heavily) overdoped
regime~\cite{timuskstatt,normanpines}. We emphasize that the
pseudogap near the antinodal region does not characterize a
precursor to the superconducting state, in the sense that the
pseudogap smoothly evolves into the superconducting gap at
$T_c$~\cite{ding,timuskstatt,normanpines}. Instead, it coexists with
the superconducting gap of the $d+s$-wave symmetry in the
superconducting state. More likely, a precursor pairing occurs in
the nodal region~\cite{leeshen}, with its onset temperature lower
than $T^*$, but above $T_c$, which may be identified with the Nernst
regime~\cite{xu,wang}. As observed, the superconducting gap near the
nodes scales as $k_B T_c$~\cite{hufner}. This makes a strong case
for our argument, if one takes into account the smallness of the
$s$-wave superconducting gap. On the other hand, ample evidence has
been accumulated, over the years, for the universal scaling relation
$E_{\rm res} = \kappa \; k_B T_c$, valid for both soft modes, i.e.,
the spin-triplet resonance mode in inelastic neutron scattering
experiments~\cite{fong1,dai,fong2,fong3,he1,he2,wilson} and the
spin-singlet mode observed as a $A_{1g}$ peak in electronic Raman
scattering~\cite{gallais,devereaux,cooper,chen,letaconetal,devereaux1},
respectively. The experimentally determined $\kappa$ is around 6,
quite close to our theoretical estimate. This presents a possible
resolution to the mysterious $A_{1g}$ problem~\cite{devereaux}: the
$A_{1g}$ mode is a charge collective mode as a bound state of
(quasiparticle) singlet pairs originating from the fluctuating $p_x
(p_y)$-wave superconducting order.

Second, is the pairing symmetry really of $d+s$-wave nature? Many
theoretical proposals, mainly based on the resonating valence bond
scenario~\cite{anderson}, predicted a pure $d$-wave
superconductivity in the
cuprates~\cite{zhangetal,gros,kotliar,suzumura}. Although a variety
of experiments have demonstrated a predominant $d$-wave
gap~\cite{scalapino,harlingen,tsuei}, strong evidence points to an
admixture of an $s$-wave component to the dominant $d$-wave
superconductivity in muon spin rotation
studies~\cite{khasanov1,khasanov2}, electronic Raman
measurements~\cite{masui}, angle-resolved electron tunneling
experiments~\cite{smilde}, and neutron crystal-field spectroscopy
experiments~\cite{furrer}. In addition, a universal scaling relation
of the superfluid density, $\rho_s(0)$, at absolute zero, with the
product of the dc conductivity $\sigma_{\rm dc} (T_{c})$, measured
at $T_{c}$, and the transition temperature $T_{c}$, indicates that a
pure $d$-wave superconductivity is realized in the
cuprates~\cite{homes}. This scaling relation may be regarded as a
modified form of the Uemura relation between the superfluid density
$\rho_s(0)$ and the transition temperature
$T_c$~\cite{uemura1,uemura2}, which works reasonably well in the
underdoped regime. However, a significant deviation from the scaling
relation was subsequently observed~\cite{tallon}, with a salient
feature that the deviation increases with doping. This feature
strongly suggests that the discrepancy should be accounted for by
removing an extra contribution from the $s$-wave component in the
context of the $d+s$-wave pairing symmetry. This issue has been
addressed recently~\cite{zhouetal}, thus supporting our scenario.

Third, any theory regarding the underlying mechanism of high
temperature superconductivity would not be complete, if the stripe
states (for reviews, see, e.g.,
Ref.~\cite{kivelson,vojta,tranquada,zaanen}) were not touched upon.
Surprisingly, a stripe-like state, i.e., a state with charge and
spin density wave order, coexisting with a spin-triplet $p_x
(p_y)-$wave superconducting state, does occur as a ground state in
the $t-J$ model for dopings up to $\delta \approx 0.18$~\cite{li},
with $J/t =0.4$. Again, all symmetries, including the four-fold
rotation and translation lattice symmetry, $SU(2)$ spin rotation and
$U(1)$ charge symmetry, are spontaneously broken. A similar line of
reasoning yields that, {\it only} the charge density wave order
survives thermal fluctuations, with the concomitant occurrence of
the soft modes: they arise from the $SU(2)$ symmetry breaking of the
spin density wave order and the spin-triplet $p_x (p_y)-$wave
superconducting order, with twice the spin-triplet $p_x (p_y)$-wave
superconducting energy gap as their characteristic energy scale.
Although it remains uncertain whether or not such a commensurate
stripe-like state is an artifact of our choice of the unit cell for
the tensor network representation of quantum states, an important
lesson we have learned from our simulation is that, the $t-J$ model
exhibits a strong tendency towards a stripe state in the underdoped
regime, consistent with the density matrix renormalization
group~\cite{white} for a more realistic stripe pattern at doping
$1/8$ and $J/t=0.35$. From this we conclude that (i) static charge
density wave order is compatible with the superconductivity, so its
possible role is deserved to be explored in the formation of the
pseudogap; (ii) static spin density wave order is detrimental to the
$d+s$-wave superconductivity, because no $d+s$-wave superconducting
component coexists with the charge and spin density order in the
ground state; (iii) fluctuating spin density wave order, together
with the fluctuating spin-triplet $p_x (p_y)-$wave superconducting
order, equally well account for the Bose-Einstein condensation in
our scenario. Therefore, the fluctuating stripe order is intrinsic
to many, if not all, families of the high temperature cuprate
superconductors.

Fourth, does our prediction about the spontaneous breaking of the
four-fold rotation symmetry and the translation symmetry under
one-site shifts in the pseudogap phase represent a physical reality?
In our scenario,  the PG phase is characterized by incoherent
preformed pairs, which occur in the anti-nodal regime in the
momentum space, leaving the remnant gapless Fermi arcs in the nodal
regime. In addition, the four-fold rotation symmetry and the
translation symmetry are broken on the square lattice. In the
superconducting phase, the four-fold rotation symmetry breaking
manifests itself in the admixture of an $s$-wave component to the
dominant $d$-wave state.  In the pseudo phase, the broken symmetry
protects the spin-singlet soft mode that is unidirectional as it
arises from the $p_x (p_y)$-wave superconducting fluctuations, as
well as the fluctuating spin density order and possibly static
charge density order. One may expect that measurable physical
effects arise from the coupling of electrons with the unidirectional
spin-singlet soft mode. Indeed, a large in-plane anisotropy of the
Nernst effect in ${\rm YBa_2Cu_3O_y}$ was reported that sets in
exactly at the pseudogap temperature $T^*$~\cite{daou}. This broken
four-fold rotation symmetry was also detected by
resistivity~\cite{r} and inelastic neutron
scattering~\cite{ns1,ns2,ns3} at low doping, and by scanning
tunneling spectroscopy~\cite{stm1,stm2} at low temperature, without
a clear connection to the pseudogap temperature $T^*$.

Fifth, what is the real bosonic glue for high temperature
superconductivity in the cuprates?  Many researchers have suggested
various candidates, such as the spin resonance
mode~\cite{demler,carbotte}, the spin fluctuation
spectrum~\cite{hayden,hwang,valla,norman,dahm}, the phonon
spectrum~\cite{lanzara,xjzhou}, a mode responsible for the dip
observed in angle-resolved photoemission spectra~\cite{davis}, and
the anti-ferromagnetic fluctuations~\cite{monthoux,millis}, as a
possible bosonic glue. However, our scenario unveils that the spin
resonance and the Raman $A_{1g}$ modes play an analogous role to a
roton in the superfluid $^4{\rm He}$, in contrast to the
conventional low-temperature superconductors with phonon-mediated
pairing. This strongly suggests that, there is no bosonic glue to
pair electrons in the cuprates, as advocated by
Anderson~\cite{anderson2006}. That is, pairing is an emergent
phenomena; it is resulted from the many-body correlations in a doped
Mott-Hubbard insulator: both the $d$-wave and the extended $s$-wave
pairings are realized to avoid strong on-site Coulomb repulsion.
Here, we remark that the spin resonance and the Raman $A_{1g}$ modes
have been suggested by Uemura~\cite{uemura2004} to be roton
analogues, by invoking soft modes in the spin and charge channels in
an incommensurate stripe state (see also Ref.~\cite{weng} for an
alternative explanation). In this sense, our scenario unlocks the
secret of the yet-to-be-achieved solid, as an analogue of the solid
$^4 {\rm He}$, in the cuprates.

In closing, let us discuss a possible candidate of the model system
that is able to reproduce all the features needed to address the
high $T_c$ problem, as mentioned at the beginning of the text. With
the fact in mind that the two-dimensional $t-J$ model does exhibit a
$d+s$-wave superconducting state and a stripe-like state in a
certain doing range for a physically realistic parameter
regime~\cite{li}, we speculate that, most likely, an extension of
the $t-J$ model to include the next-nearest-neighbor and the
third-nearest-neighbor hopping terms, which are known to be
significant in the cuprates, will do the job.

We thank Sam Young Cho, Sheng-Hao Li, Bo Li, Qian-Qian Shi, Hong-Lei
Wang, Zu-Jian Ying and Jian-Hui Zhao for useful and stimulating
discussions. This work is supported in part by the National Natural
Science Foundation of China (Grant Nos: 10774197 and 10874252).

\end{document}